\documentclass[review]{elsarticle}
\usepackage[bottom=2.0cm,top=2.0cm,left=1.7cm,right=1.7cm]{geometry}
\usepackage{epsfig}
\usepackage{amssymb}
\usepackage{amsmath}
\usepackage[utf8]{inputenc}
\usepackage[T1]{fontenc}
\usepackage{graphicx}
\usepackage{caption}
\usepackage{subcaption}
\usepackage{textcomp}

\begin{document}
\begin{frontmatter}
\title{The black hole of logistics costs of digitizing commodity money}
\author{Boliang Lin$^{abc}$ $^\ast $, Ruixi Lin$^d$}
\address{$^{a}$ School of Traffic and Transportation, Beijing Jiaotong University, Beijing 100044, China \\
 $^{b} $ Sino-UK Blockchain Industry Research Institute, Guangxi University; Nanning 530004, China\\
  $^{c} $ Beijing Laboratory of National Economic Security Early-warning Engineering, Beijing 100044, China\\
   $^{d}$ School of Computing, National University of Singapore, Singapore, 117416, Singapore \let\thefootnote\relax\footnotetext{
  $^{*}$Corresponding autohr. Email:bllin@bjtu.edu.cn}}


\begin{abstract}
In this paper, we reveal the depreciation mechanism of representative money (banknotes) from the perspective of logistics warehousing costs. Although it has long been the dream of economists to stabilize the buying power of the monetary units, the goal we have honest money always broken since the central bank depreciate the currency without limit. From the point of view of modern logistics, the key functions of money are the store of value and low logistics (circulation and warehouse) cost. Although commodity money (such as gold and silver) has the advantages of a wealth store, its disadvantage is the high logistics cost. In comparison to commodity money, credit currency and digital currency cannot protect wealth from loss over a long period while their logistics costs are negligible. We proved that there is not such honest money from the perspective of logistics costs, which is both the store of value like precious metal and without logistics costs in circulation like digital currency. The reason hidden in the back of the depreciation of banknotes is the black hole of storage charge of the anchor overtime after digitizing commodity money. Accordingly, it is not difficult to infer the inevitable collapse of the Bretton woods system. Therefore, we introduce a brand-new currency named honest devalued stable-coin and built a attenuation model of intrinsic value of the honest money based on the change mechanism of storage cost of anchor assets, like gold, which will lay the theoretical foundation for a stable monetary system.
\end{abstract}


\end{frontmatter}

\section{\label{Introduction}Introduction}


Without sound money, there is no protection for savings and property, nor capital accumulation, nor long-term investment, nor entrepreneurship, nor social advance. Although it has long been the dream of economists to stabilize the buying power of the monetary units, the goal we have honest money always broken since the central bank depreciate the currency without limit( Ebeling, 2015)$^{[1]}$. The reason hidden in the back of the depreciation is the black hole of storage charge of anchor over time.

Generally, commodity is a labor product that is used for exchange. Obviously, the basic property of commodity is values and use values. Therefore, commodity also has the property of finance. According to the common definition, commodity money is money whose value comes from a commodity of which it is made. i.e., commodity money consists of objects that have value in themselves as well as value in their use as money (Sullivan et al., 2003)$^{[2]}$. It mainly includes physical money and metal money in money-shape. Physical money whose value as money is equal to its value as a commodity is the primal money-form produced in the long-term development process of commodity exchange. At the early stage of human society, it was fixed on some specific kinds of commodities, for example, cattle and sheep, shells and so on.


From the perspective of logistics, the transportation and storage of commodity will incur logistics costs. Although the physical money has an advantage of stable intrinsic value, it has an obvious drawback of high logistics cost. In comparison, the logistics cost of metal money is much lower. Therefore, metal money gradually replace other physical money in the process of the development of human beings.

In order to further reduce the logistics cost of physical money, there was a time when people used precious metals as a major commodity money. However, due to the limited resources of precious metals, the growth of precious metals is far slower than the accumulation of human wealth. In addition, people's expectations that precious metals will appreciate often lead to the behaviors of hoarding. This kind of phenomenon will inevitably result in liquidity shortage in the money market.


The emergence of representative money is a milestone to reduce logistics cost of commodity money. Generally, representative money, as the receipt for circulated metal money, is those paper money or banknotes issued by the government or banks, which replaces metal money in circulation. In theory, the representative money is equivalent to commodity money. In fact, the value of representative money is often lower than its anchor value represented by the goods. The hidden reason is to be discussed later in this paper. Compared with metal money, representative money not only has the advantages of low printing cost and low logistics cost, but also can overcome the weight loss caused by metal money in circulation.


The evolution of the representative money is credit currency, which completely bid farewell to the logistics cost incurred by commodity money. But the inherent defect of credit currency is the risk of unrestrained over-issue. In recent years, people are anxious to inflation of fiat money and mistrust of the central bank.


Hence, some decentralized digital currencies, for example, Bitcoin, have emerged. Actually, the virtual currencies do not correspond to real wealth since their supply is only controlled by algorithms. Virtual currencies are a type of digital currency, typically controlled by its creators and used and accepted among the members of a specific virtual community (Rastogi et al., 2019)$^{[3]}$. Moreover, their prices have fluctuated significantly in recent years. Stablecoin is an approach to address this problem of high price volatility, which is for example defined as a digital currency that is pegged to another stable asset like gold, or to major fiat currencies like Euros, Pounds or the US dollar (Mita et al, 2015)$^{[4]}$.


However, all these money with zero logistics cost, namely modern money, is basically token money. Therefore, it is irrelevant to its intrinsic value and not a currency of "stable value". In this background, those credit currency holders will be the victims of inflation. Therefore, it is an important research topic on how to reveal formation mechanism of the cost of symbolized or certificated commodity money from the perspective of logistics costs, and how to design a new currency of "stable value" to integrate the advantages that credit currency has no logistics cost and that commodity money can store wealth.


This paper aims to reveal the depreciation mechanism of representative money (banknotes) from the perspective of logistics warehousing costs. The contributions of this paper can be summarized as follows:\\
$\bullet$ We proved that there is not such honest money from the perspective of logistics costs, which is both the store of value like precious metal and without logistics costs in circulation like digital currency.\\
$\bullet$ A brand-new currency named honest devalued stable-coin is introduced.\\
$\bullet$ A attenuation model of intrinsic value of the honest money based on the change mechanism of storage cost of anchor assets is built.

The remainder of this paper is organized as follows. Section 2, we review the evolution of money form from the perspective of logistics cost. In Section 3, a technique route for a currency of stable value is described. Section 4, from the perspective of a goldsmith is the warehouse night watchman, we analyzed the phenomenon which inventory cost of mortgage assets were intentionally hide in the process of signifying commodity money. In Section 5, we put forward the paradox of the existence of a currency of "stable value". A honest currency of “stable devalued” model based on “gold” standard is proposed in Section 6. Finally, conclusions are drawn and future research directions are discussed in Section 7.


\section{\label{}Review the evolution of money from the perspective of logistics cost}

Early trade took the form of barter that goods are directly exchanged for other goods without using a medium of exchange. For example, a sheep can be exchanged for a stone axe. However, the goods exchanged sometimes are not what the other party needs. Therefore, it is necessary to find an object that can be acceptable to both in exchange, which is the primitive money. As a result, some items that are not easy to be obtained a large quantities, such as rare shells, fabrics, feathers of rare birds, precious stones, alluvial gold, etc., are assumed to act as the universal equivalent, namely primitive commodity money, at different times in different regions. On average, the logistics cost of primitive commodity money is lower than barter. However, this kind of equivalent commodity still has fatal disadvantages such as difficulty in carrying and storing. That is to say, it has high logistics cost.

Universal equivalent, as the medium of exchange, needs to frequently circulate among commodity producers. Obviously, it will be touched by countless people. What’s more, money usually can be saved to store value for a long time, which retains purchasing power into the future. However, it is impractical for some commodity money, such as shells, fabrics, cocoa beans, to store wealth for the future.

 With the second great division of labor in society, handicraft industry was separated from agriculture. Due to the inventions in melting technology, metal gradually become the mainstream of universal equivalent in trading, which made metals as coin materials. Generally, metals, such as gold, silver and copper etc., have the advantages of stable value, divisibility and easy to carry. As a result, they played the role of money in a long time. Obviously, the logistics cost of metal money has been reduced significantly.


In the early days of commodity economy, most countries and regions used base metals as money, for example, copper. Base metals are compatible with the initial commodity economy. However, there was a problem later that monetary materials competed for raw materials to used for manufacturing. Moreover, base metals lost their suitability as media of exchange when their value per pound became too low. For example, the copper money of seventeenth-century in Sweden was notoriously cumbersome. Individual pieces of copper "plate money" eventually weighed up to 20 kilograms (44 pounds). Strong young men had to be employed to carry the copper to make an ordinary-sized commercial purchase, which made the logistics cost extremely high. Finally, Swedes stopped using copper, except in some small trading (White, 1999)$^{[5]}$. In the 19th century, many countries in the world joined the Latin Monetary Union (LMU) and adhered to bimetallism (Bae et al., 2011)$^{[6]}$.


 Paper money emerged after metal money, also known as credit currency, has almost no logistics cost. Actually, paper money is a promise or liability which represents physical assets by a piece of paper that circulates as medium of exchange (Battilossi et al, 2018)$^{[7]}$. Of course, in a serious inflation period, people need to use a lot of money to buy daily necessities. When the value of paper money is similar to its cost of production, there is still high logistics cost.

 After entering the Internet era, electronic money has been widely used. Electronic money will inevitably reduce the time and space disposal of expenses of payment and settlement transactions, which increases the volume of transactions by promoting transaction convenience (Mohamad et al., 2002)$^{[8]}$. In this case, people can realize the long-distance transmission and payment of money in a short time which makes the transactions convenient. Therefore, the circulation of money has truly entered the era of no logistics cost.

The logistics cost of money in various historical periods are shown in Fig.1. Among them, cattle and sheep are the representative of barter, and shells are the representative of the primitive commodity money.

\begin{figure}[h]
\vspace{2mm}\begin{center}
\includegraphics[scale=1]{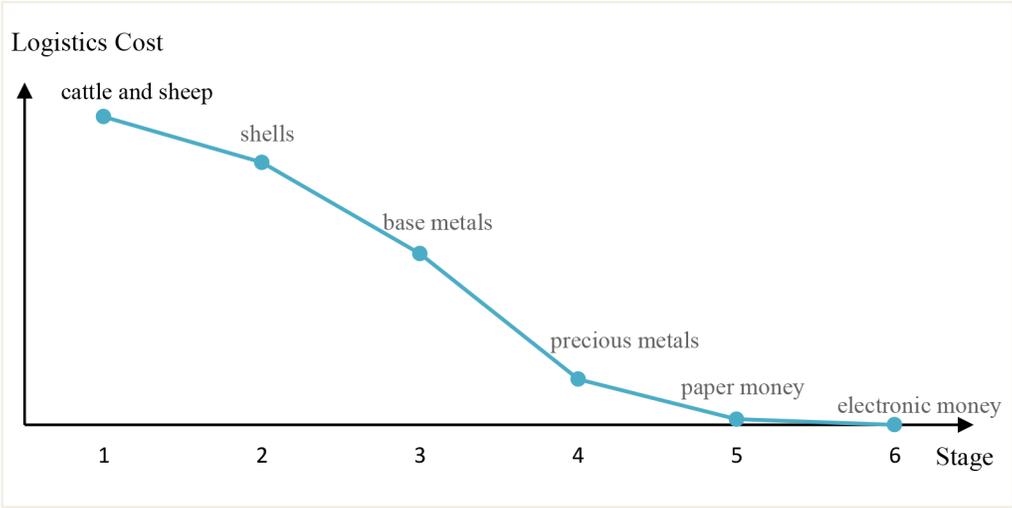}\\
\caption{Logistics cost for each stage of monetary development}
\label{Fig 1.}
\end{center}\vspace{2mm}
\end{figure}

 In the process of monetary evolution, human beings always tend to choose money with lower logistics cost. Obviously, as a medium of exchange, the minimization of logistics cost is one of the goals pursued in the process of monetary development.

\section{\label{}A technology framework to design a currency of stable value based on the trade-offs between store of value and storage cost of anchor}

The first function of money is measure of values. Just as a weight used on a balance itself has weight, obviously, money used to measure the value of a commodity itself is also a commodity and has intrinsic value. Therefore, commodity money can serve as a measure of value.

 After the seller converts goods into money in the trading process, the money will be deposited for a period of time, and then be used to buy other goods. In these processes, there will be logistics costs for carrying and storing metal money. Therefore, in order to reduce the cost in exchange of goods and money, it is not difficult to understand that people choose the precious metal material with the highest intrinsic value per unit weight or volume to cast specie. In fact, it is the embodiment of minimizing the logistics cost of commodity money.

 On the one hand, specie is constantly being worn out in circulation, on the other hand, considering that the function of money as a medium of circulation is only a transient medium. Therefore, either devalued specie or even a currency symbol with no intrinsic value at all, can also replace metal money in circulation. From this point of view, to symbolize commodity money is the best way to prevent it from being worn and tore in circulation.

 As a store of value and a universal representative of asset or material wealth, money should have the function to be directly convertible into any other commodity. In other words, it must be full-bodied commodity money, such as gold and silver specie or bullion. Generally, only full-bodied metal money, namely honest money that its value is honestly established in the marketplace without bankers, government, politicians, or the Federal Reserve manipulating its value to serve special interests (Paul, 2008)$^{[9]}$, can be saved and serves as a store of value. On the contrary, it is not stable for the wealth represented by hoarding paper money. Obviously, storage cost of paper money is much lower than that of the metal money. However, the devaluation rate of credit currency is unpredictable, which renders it hard to store wealth. It is a basic fact that since credit currency dominated the Global Financial System, Western developed countries have been fighting against inflation, and the harm caused by hyperinflation has been obvious to all.

 In a nutshell, honest money (physical money) is the best choice considering the functions of a measure of value and a store of value. Nonetheless, the main drawbacks are as follows: first, logistics cost is high in circulation; second, the weight of the metal loss is not easy to be controlled; third, it is hard to avoid hoarding of money, especially precious metals. From the perspective of a medium of circulation, paper money and digital money with zero logistics cost are the best choices.

 Based on what has been discussed above, from the perspective of logistics cost, the design approach of future new money should satisfy the following constraint conditions: (1) it has almost no logistics cost including transportation or carrying cost in circulation and safekeeping cost in storage; (2) It has no weight loss in circulation; (3) It has the function of storing wealth; (4) Hoarding of money can be avoided which will not result in a shortage of market liquidity;(5) It contains the property of scarcity resource, such as precious metal money (commodity money), and over-issue is not easy.

 The above characteristics of the new currency actually are a combination of physical money and credit currency. According to these analyses, the best design approach for future money is the digitization of physical money.

\section{\label{}Inventory cost of mortgage assets were intentionally hide in the process of signifying commodity money}

About a thousand years ago, base metal money, mainly iron and copper coins, was circulated as currency in Sichuan, China. When merchants faced with large transactions, carrying heavy metal money will significantly increase logistics costs. To reduce the carrying cost of specie in circulation, some businessmen set up deposit shops named “jiaozi” shops to specialize in safekeeping of metal money for merchants with large sums of money. In this case, the depositors hands over their money to the shop, and then were given a slip of special paper named “jiaozi” that made from mulberry leaves, recording the amount of cash they had with that shop. If the “jiaozi” holders show the paper to that shop, they could redeem on demand in metal money after paying a safekeeping fee of 3{\%}  (Xue, 2015)$^{[10]}$. Actually, “jiaozi” is regarded as the first paper money used in China and even in the world. Afterwards, many merchants jointly established the deposit shops specializing in the issuance and exchange of “jiaozi”, and set up branches in other places. Since the deposit shops abide by their credit and keep the promise, “jiaozi” could be converted into money at any time. Eventually, “jiaozi” were served as currency for decades in the region (Battilossi et al., 2018)$^{[7]}$.

 In the middle of the 17th century, a lot of goldsmiths engaged in the production of gold jewelry. Some civilians got rich through business, and they put their gold stored in the goldsmith’s shops, then goldsmith wrote a receipt for the depositors. Depositors can withdraw their gold at any time with the receipt, at that time, goldsmiths needed to be paid safekeeping fees and handling fees. The further development of the business, goldsmiths can also transfer gold and silver from the depositor A to the customer B upon the written request of depositor A. From the perspective of modern logistics, the early goldsmith in England is equivalent to the current third-party logistics, and the goldsmiths were the leading financiers in England (Milevsky, 2017)$^{[11]}$.

 Later, the receipt of safekeeping gold coins evolved into a format certificate. The depositors or certificate holders found that when they needed money to buy goods, they did not have to withdraw the gold at all and just gave the certificate to the supplier. And then later, goldsmiths realized that the certificates they had issued could be used as money. Therefore, they began to issue “fake certificates” for lending and reaped the benefits. As long as all depositors do not come to withdraw gold on the same day, “fake certificates” are equivalent to “real certificates”. This shows the origin of the “deposit reserve system” in modern banks. In a nutshell, physical money was replaced by “fake certificates”, and the goldsmiths, as third-party logistics, evolved into banks. So, goldsmiths not only are no longer charged the storage fee but also paid some fees to depositors. Actually, this method was adopted by ancient Chinese banks, generally called qianzhuang, whose certificates were called “yinpiao”, a bank notes or receipt of storage silver (Yang et al, 2019)$^{[12]}$. In this way, the safekeeping fee of gold and silver magically disappear in the black hole of the certificate market.

 As the safekeeping receipts evolved into bank notes, transferring certificates evolved into bank checks, and full reserves turned into fractional reserves system, the goldsmith industry gradually developed into a banking industry with main business on money. The essential practice of modern banks is still to lend out the depositors’ assets to obtain profits, which will cover safekeeping fee (inventory costs) and interest expenditure for depositors.

 From the above analysis on the origin of banks and credit money, we can find that as an honest money with the function of a store of wealth must be a commodity money, and any bank note will enter a devaluation channel if they can’t redeem on demand in gold or other assets of anchor. In order not to reduce the monetary function of a store of wealth, only to restore the goldsmith as the warehouse night watchman. In this way, we can put the logistics cost of commodity money back into people’s sight from the disappearing black hole. And this cost would be clearly reflected on the bank notes or certificates of anchor goods.

\section{\label{}The paradox of the existence of a currency of "stable value"}

Stablecoin is a cryptocurrency that can keep a stable exchange rate with its anchor. It is designed to be price stable with respect to some anchor (Moin et al., 2020)$^{[14]}$.


The stablecoin can be divided into the following several types according to the different anchors. (1) Fiat Stablecoins, backed by fiat money as collateral; (2) Digital stablecoins, backed by crypto-assets like Bitcoin as collateral; (3) Commodity Stablecoins, backed by commodities like precious metal as collateral; (4) Algorithmic Stablecoins without collateral, backed by users’ expectations about the future purchasing power of their holding.

Among them, commodity stablecoin is the most expected choice. In theory, the anchor goods of stablecoins can be precious metals, steel, real estate, etc. For instance, AnthemGold, a Texas-based blockchain company, launched its gold-backed asset token AGLD (Business Wire, 2019)$^{[14]}$. Each token is backed by one gram of gold. Evidently, it is a stablecoin that anchors physical assets.

Up to now, there is no strict definition of stablecoins. In the following text, we will define a stablecoin as one kind of certificate that can be converted into mortgage assets at any time. It can be considered as a currency of "stable value". Next, we will prove that it is impossible for the currency to exist.

 We assume that the customer $k$  buys $n_k$  tokens of AGLD at time $t_k$ with $n_k$ grams of gold (or equivalent fiat money, like US dollars at the same time and place). After a period of time, the customer or the person who is in possession of tokens redeems gold at time $t_{k}^{\prime}$ . The customer will pay the processing fee to the AnthemGold company is $\beta$ per token, thus the gross profit issuer can obtain is formulated as follows:\\
\begin{equation}
\label{eq1}
C^{\mbox{profit }}=\sum_{k} \beta \times n_{k}
\end{equation}
\indent The logistic cost of the issuer can be formulated as follows:\\
\begin{equation}
\label{eq2}
C^{\mbox {warehouse }}=\sum_{k} \alpha \times n_{k} \times\left(t_{k}^{\prime}-t_{k}\right)
\end{equation}

Where, $\alpha$  is warehouse cost of saving one gram of gold per day, which consists of transportation cost, settlement cost, storage cost, management cost, etc.\\ \indent If the gross profit of issuer can't cover logistic cost, i.e., the follow condition will be met,\\
\begin{equation}
\label{eq3}
C^{\mbox {profit }}<C^{\mbox {warehouse }}
\end{equation}

then the issuer will go bankrupt. Obviously, there exist always such a day when condition (3) is met. Without loss of generality, assuming that the whole system has only one customer, formula (3) can be simplified as $\beta \times n<\alpha \times n \times\left(t^{\prime}-t\right)$. We can still find the buying time $t$  and selling time $t'$ of the stablecoin AGLD, which makes the following formula (4) work.\\
\begin{equation}
\label{eq4}
t^{\prime}-t>\frac{\beta}{\alpha}
\end{equation}

 In order not to go bankrupt, after collecting the user's collateral, the issuer invests the collateral during the user holds the stablecoin, which will earn interest for issuer. In this case, investment risk will be generated. Therefore, stablecoins will be unstable. AGLD is similar to the U.S dollar during the Bretton Woods system, except that there was no function to convert gold to individual holders of US dollars. The Bretton Woods system priced an ounce of gold at {\$}35.2, therein {\$}0.2 was for transportation fee. According to formula (4), the accumulation of its warehouse cost will eventually consume all the gold. Hence, the Bretton Woods system is bound to collapse, which means that the decoupling of the dollar from gold is a historical necessity.

  In the same way, "a 100 Percent Gold Dollar" proposed by Rothbard$^{[15]}$ in the last century cannot exist stably, because he did not propose a gold redemption route that face value devalues over time.

 In conclusion, for any currency anchored by 100{\%} commodities, as long as it does not have a timestamp and its face value does not devalue over time, then the currency cannot become a stablecoin or as a currency of "stable value".

\section{\label{}A honest currency of "stable devalued" model based on "gold" standard}

We have proven that an absolute currency of stable value is impossible to become a reality, except for commodity money. Although the logistical costs of digital currency and banknotes can be negligible, they hardly fulfill the function of money as a store of value. While commodity money such as precious metals can store wealth, the logistical costs of circulation are too high. Taking the advantages of both, we innovatively design an honest currency of “stable devalued”, a brand-new currency named (DSC, similar to DCM in reference [16]) and built a attenuation model of intrinsic value of the honest money based on storage cost of anchor assets. Obviously, a SDC issuer must adopt full reserve system, and the anchor of the SDC should be some physical commodities rather than a credit currency.

\subsection{\label{}The Design Principle of An honest stable devalued coin}
Supposing a currency issuing institution (a bank or a finance company) $B$  issues a type of  SDC at time  $t$ (usually in the New Year's day of a certain year such as January 1, 2035). Each SDC contains a gram of precious metal $m$  as a mortgage assets. Issue number is $N_{B}^{m}(t)$. Let  $M_{B}^{S D C}(t, m, W, \rho, \theta, E, \Delta W)$ denote the certificate of a SDC, where $t$ is the date of issue, and $m$  stands for the mortgage assets, $W$ for the weight of anchor goods each SDC, $\rho$ for purity of the the metal materials as anchor assets, $\theta$  for the attenuation coefficient of the weight of anchor goods, $E$ for the expiry date of SDC and  $\lambda$ is the delivery fee rate for SDC holder to redeem for collateral. Take gold as a case, for example, $m$  is gold, $W=1$ , $\rho=0.9999$ , $E$ is fifty years, and the volume of issuing is 2 billion shares (2,000 tons) , $\Delta W=3 \times w^{\mbox{remaining }}\left(t^{\mbox {redemption }}\right) / 1000$ ,  $t^{\mbox {redemption }}$  is the redemption time and $w^{\mbox {remaining }}\left(t^{\mbox {remaining }}\right)$  is the residual weight of a SDC. The redemption fee equals to 3\textperthousand  of the weight of the residual value, i.e., $\lambda$= 3\textperthousand  .


 Thus, the residual weight of a SDC at date $t+\Delta t^{\prime}$   is\\
\begin{equation}
\label{eq5}
W\left(t+\Delta t^{\prime}\right)=W \times \theta^{\Delta t^{\prime}}
\end{equation}

\indent Let  $P_{B}^{m}\left(t+\Delta t^{\prime}\right)$ denote the price of the precious metal $m$ with purity $\rho$ , where the price is set according to the local currency. Then, a customer needs to pay  $P_{B}^{m}\left(t+\Delta t^{\prime}\right) \times W \times \theta^{\Delta t^{\prime}}$ for a SDC. In practice, the customer can give the issuer $W\left(t+\Delta t^{\prime}\right)$ units of the precious metal $m$ for a SDC. In this case, the issuing institution $B$ will inspect whether the precious metal submitted by the customer meets the purity standards, and this will incur an inspection fee of $\Delta H$. Considering that the local currency is not stable, $\Delta H$ could differ at different times, and the issuing institution $B$ will have to announce $\Delta H$ in advance. 

 When a SDC holder needs to redeem the collateral of a SDC at time $t+\Delta t^{\prime \prime}$ , the quantity of precious metal available to the customer is\\
\begin{equation}
\label{eq6}
W_{m B}^{\mbox{Redemption }}\left(t+t^{\prime \prime}\right)=\theta^{t^{\prime \prime}} \times W-\lambda \times \theta^{t^{\prime \prime}} \times W=(1-\lambda) \theta^{t^{\prime \prime}} W
\end{equation}

The second item in the formula (6) is fees arising from such delivery at redemption.  

 Of course, the issuer can require that the amount of the precious metal $m$ redeemed must be an integer multiple of certain units (for example, one kilogram of gold is the minimum unit). In any case, SDC holders have a relatively clear expectation of the future wealth.
\subsection{\label{}A Numerical Examples of SDC}
 To further illustrate the innovative principle of SDC, we design an honest stable devalued coin named SDC\_AU999, issued by LIN on January 1, 2030, with a circulation of one billion coins. Each coin is backed by one-gram of gold with a fineness of 99.9{\%}. As mortgage assets, one thousand tons of gold were entrusted to a warehousing company in Switzerland to be stored in caves in the Alps. Besides, the marking information for each SDC\_AU999 includes : the issuing date is on January 1, 2030, the mortgage asset is 99.9{\%} gold, the weight of associated asset is 1 gram, the daily attenuation coefficient is  $\theta=99.996 \%$, the delivery fee is 3\textperthousand of the actual delivery weight, the minimum delivery weight is 1000g and the delivery point is appointed by LIN. and the validity is 50 years, i.e. December 31, 2079 (see notes for redemptions beyond validity). 

 When a client want to buy 2000 coins, i.e., certificate of SDC\_AU999, on July 1, 2030 in New York, the actual weight of mortgage assets for each coin after 183 days will be:\\
\begin{equation}
\label{eq7}
W(183)=W \times \theta^{183}=1 \times 0.99996^{183}=0.9927(\mathrm{~g})
\end{equation}
\indent If the global gold price on that day is {\$}1500 /OZ and the selling price in New York is 101{\%} of that, the money that the client needs to pay will be:\\
\begin{equation}
\label{eq8}
\begin{aligned}
P_{m B}^{\mathrm{Sell}}(183) &=(1500 \times 1.01) \times W \times \theta^{183} \\
&=1515 \times 0.9927 / 31.1035=48.35(\mathrm{USD})
\end{aligned}
\end{equation}

 Thus, the client should pay 96705.55 USD for 2000 coins of SDC\_AU999 actually. 

 The opposite is the SDC holders to redeem the mortgage assets. When a SDC holder wants to redeem for gold on January 1, 2031, the residual weight of 2000 coins of SDC\_AU999 will be:\\
\begin{equation}
\label{eq9}
W(365)=2000 \times 0.99996^{365} \times(1-0.003)=1965.0986(\mathrm{~g})
\end{equation}

 Considering that the minimum unit of delivery weight is 1000g, the client can make up for the payment for goods according to the gold price of the day, and then can get 2000 grams of gold with purity of 99.9{\%}.

 If the client wants to convert SDC into the currency of the country, one way is to sell DCM-SDC on the market; and another is to sell SDC back to issuer. Assuming that the repo rate of the issuer recycle a SDC is 2\textperthousand of the residuary weight, and then the chargeable weight will be:
\begin{equation}
\label{eq10}
W(365)=2000 \times 0.99996^{365} \times(1-0.002)=1967.0698(\mathrm{~g})
\end{equation}

 Assume the gold price is {\$}1600 / OZ on that day, the charge that issuer should pay to the client will be:
\begin{equation}
\label{eq11}
P_{m B}^{\mbox {buy-back }}(365)=1967.0698 \times 1600 / 31.1035=101188.36 (\mathrm{USD})
\end{equation}

 The mode of SDC circulation can be similar to the paper money or digital currency, or to Bitcoin's mode which is through an open management account transfer mechanisms. In order to prevent SDC over issue by some authorized banks, the SDC’s mortgage assets which are stored by third-party logistics service provider need to be over sighted by the third-party needs to receive a third party must be supervised by relevant international organizations like IMF.

Of course, banks can also accept SDC deposits, similar to the rules for fiat deposits. Obviously, the devalued stablecoin is a good money that can store wealth well. Moreover, due to the SDC own attenuation function, it cannot be widely hoarded by people, which can avoid the risk of a shortage of liquidity.
\subsection{\label{}The Model for determining the attenuation coefficient of SDC}
 For example, the financial institution $B$  issues $N$  certificates of SDC anchored with precious metal $m$ at the time of $t$. If the initial weight of each SDC is $W$, the total initial weight of mortgage assets is $N \times W$ . Assuming the validity period of the SDC is $E$, we can split $E$  into $D$ intervals $\{1,2, \cdots, d, \cdots, D\}$. Let  $w^{\mbox {remaining }}(d)$ denote the residual weight and $\Delta w(d)$ is the extracted weight from a SDC in the intervals $d$ (at the end of the period of time), we have the expression as follows:
\begin{equation}
\label{eq12}
\Delta w(d)=w^{\mbox {remaining }}(d)-w^{\mbox {remaining }}(d-1)=\mu(d) \times w^{\mbox {remaining }}(d-1)
\end{equation}

where $\mu(d)$ is the attenuation coefficient of SDC in the intervals $d$.

 Let $c^{\mbox {storage }}(d)$  denotes the storage charge to store a unit weight of the metal $m$ in the intervals $d$, which including warehouse rent, management staff salary, etc., and $P _{m}^{\mbox {precious\_metal }}(d)$ denotes the price of the precious metal  $m$ at that time, the condition that the issuer does not lose money is:\\
\begin{equation}
\label{eq13}
P_{m}^{\mbox {precious\_metal }}(d) \times \Delta w(d) \geq c^{\mbox {storage }}(d) \times w^{\mbox {remaining }}(d-1)
\end{equation}
\begin{equation}
\label{eq14}
\qquad P_{m}^{\mbox {precious }_{\mbox {metal }}}(d) \times \mu(d) \geq c^{\mbox {storage }}(d)
\end{equation}

 So, we can get the condition:\\
\begin{equation}
\label{eq15}
\mu(d) \geq \frac{c^{\mbox {storage }}(d)}{P_{m}^{\mbox {precious\_metal }}(d)}
\end{equation}

A simple way to calculate the attenuation coefficient is to take the average of all $\mu(d)$ :\\
\begin{equation}
\label{eq16}
\bar{\theta}=\frac{1}{D} \sum_{d=1}^{D} \mu(d)=\frac{1}{D} \sum_{d=1}^{D} \frac{c^{\mbox {storage }}(d)}{P_{m}^{\mbox {precious\_metal }}(d)}
\end{equation}

 It can be seen that the attenuation coefficient depends on the predicted price and storage costs of the precious metal $m$  for every intervals of time.

  During the expiry date  $E$, some SDC holders may redeem their mortgage assets. The issuer can resold these SDCs or destroy it.

 In fact, there are always some SDCs backlog in the issuer's hand, which are in a state of for sale and will incur the cost of capital tied up. Therefore, the attenuation coefficient need to have a small correction as follow:
\begin{equation}
\label{eq17}
\theta=\bar{\theta}+\Delta \theta
\end{equation}

 The size of parameter $\Delta \theta$ should be determined by issuer according to the estimate speed of the SDC sales.
\subsection{\label{}An approach to redemption of SDC over expiry date}
If the SDC holder still does not redeem the mortgage assets of SDC from the issuer $B$ in expiry date $E$ with the version, the issuer shall have the right to set new attenuation coefficient for the expired certificates.

 We can cope with this problem in two different ways in regards to the conditions. First, if the issuer $B$ has recently issued a version of  SDC' before the expiry date $E$ and its expiry date $E^{\prime}>E$, the default attenuation coefficient is equal to version SDC'. The second case is the above condition is not met. In this case, the issuer $B$ need to refer the attenuation coefficient which of the similar SDC issued by other issuers.

 Supposing that there are $N^{\mbox {bank }}$ banks issued the SDC with precious metal  $m$ as mortgage assets during the period of time $(E-\Delta E, E)$, the corresponding attenuation coefficients are  $\left\{\theta_{1}, \quad \theta_{2}, \cdots, \quad \theta_{N_{m}^{b a n k}}\right\}$ respectively. The mean value $\bar{\theta}$ is:\\
\begin{equation}
\label{eq18}
\bar{\theta}=\frac{1}{N_{m}^{b a n k}} \sum_{k=1}^{N_{m}^{b a n k}} \theta_{k}
\end{equation}

 The maximum one of these attenuation coefficients is as follows:\\
\begin{equation}
\label{eq19}
\theta^{\mathrm{max}}=\max \left\{\theta_{1}, \quad \theta_{2}, \cdots, \quad \theta_{N_{m}^{b a n k}}\right\}
\end{equation}

 Considering that SDC holders should  the blame bear since they did not redeemed within the prescribed time limit, the determination of the attenuation coefficient of some overdue SDC can be higher than the average value of the market as follow:\\
\begin{equation}
\label{eq20}
\theta^{\mbox {overdue }}=\bar{\theta}+\alpha \times \theta^{\max }
\end{equation}

where the coefficient  can take a certain value of $\alpha \in\{0,0.5\}$.
\section{\label{}Conclusions}
Strictly speaking, for any currency anchored by 100{ \%} commodities, as long as it does not have a timestamp and its face value does not devalue over time, then the currency cannot become a stablecoin or as a currency of "stable value". Although commodity money itself, such as precious metal money, has its inherent basic value, a major drawback of physical money is high logistics costs in circulation. When representative money comes into use, such as silver certificate issued by private financial institutions in Shanxi one or two hundred years ago, in theory, the logistics cost reduces to nearly zero. However, this is based on the assumption that banks, as the third-party logistics, provide customers with physical currency custody and exchange it for physical goods without compensation. In fact, we cannot assume that free storage always exists. Hence, financial institutions that keep anchor goods of representative money generally lack the integrity owned by third-party logistics companies. They would inevitably appropriate anchor goods for other uses or set up obstacles to conversion, such as inconvertibility, to gain profits and pay storage fees. In history, the cost of safekeeping anchor goods is basically realized through the risk transfer of lending, while credit money transfers its risks through the devaluation. Therefore, there is no absolute stablecoin whose value does not decay in proportion to time.

Due to the progress of technology and the cost of access to resources in the fall, it is difficult to try to stabilize the buying absolutely power of a monetary units. One of the feasible ways to build an honest money system is to symbolize commodity money as a certificate which can redeem mortgage assets on demand. A difficult problem to solve is how to pay issuer who is as the warehouse night watchman the safekeeping fee. How fast is a rational rate for devaluation of denomination of certificate, which is an acceptable solution to both the issuer and holders of certificate? In addition, in consideration of the dramatical price fluctuation of the current virtual currencies, such as Bitcoin, we also can set an attenuation coefficient for them in order to increase stability.

\section*{\label{}Acknowledgments}
This research is supported by the Program of the Co-Construction with Beijing Municipal Commission of Education of China (Grant No. B20H100020, B19H100010). The authors wish to thank Prof. Ning Wang of Mathematical Institute, University of Oxford , Prof. Xiaojun Jia of Beijing Laboratory of National Economic Security Early-warning Engineering and Mr. Xiaohong Zhao of Beijing Jiaotong University for their valuable comments and suggestions.


\end{document}